\documentclass[aps, prl, reprint, superscriptaddress]{revtex4-2}
\pdfoutput=1
\usepackage{braket,dsfont,subfigure,amsfonts,amssymb,bm,graphicx,float,amsmath,amsthm,txfonts}
\usepackage[colorlinks]{hyperref}
\usepackage{enumitem}
\usepackage{orcidlink}
\newcommand{\Tr}{{\rm Tr}} 
\newcommand{\llvert}{\left|\left|}
\newcommand{\rrvert}{\right|\right|}

\theoremstyle{definition}

\begin{document}
\title{Efficient Quantum Circuit Compilation for Near-Term Quantum Advantage}
\author{Yuchen Guo~\orcidlink{0000-0002-4901-2737}}
\affiliation{State Key Laboratory of Low Dimensional Quantum Physics and Department of Physics, Tsinghua University, Beijing 100084, China}
\author{Shuo Yang~\orcidlink{0000-0001-9733-8566}}
\email{shuoyang@tsinghua.edu.cn}
\affiliation{State Key Laboratory of Low Dimensional Quantum Physics and Department of Physics, Tsinghua University, Beijing 100084, China}
\affiliation{Frontier Science Center for Quantum Information, Beijing 100084, China}
\affiliation{Hefei National Laboratory, Hefei 230088, China}

\begin{abstract}
    Quantum noise in real-world devices poses a significant challenge in achieving practical quantum advantage, since accurately compiled and executed circuits are typically deep and highly susceptible to decoherence.
    To facilitate the implementation of complex quantum algorithms on noisy hardware, we propose an approximate method for compiling target quantum circuits into brick-wall layouts.
    This new circuit design consists of two-qubit CNOT gates that can be directly implemented on real quantum computers, in conjunction with optimized one-qubit gates, to approximate the essential dynamics of the original circuit while significantly reducing its depth.
    Our approach is evaluated through numerical simulations of time-evolution circuits for the critical Ising model, quantum Fourier transformation, and Haar-random quantum circuits, as well as experiments on IBM quantum platforms.
    By accounting for compilation error and circuit noise, we demonstrate that time evolution and quantum Fourier transformation circuits achieve high compression rates, while random quantum circuits are less compressible.
    The degree of compression is related to the rate of entanglement accumulation in the target circuit.
    In particular, experiments on IBM platforms achieve a compression rate of $12.5$ for $N=12$, significantly extending the application of current quantum devices.
    Furthermore, large-scale numerical simulations for system sizes up to $N=30$ reveal that the optimal depth $d_{\rm max}$ to achieve maximal overall fidelity is independent of system size $N$, suggesting the scalability of our method for large quantum devices in terms of quantum resources.
\end{abstract}

\maketitle

\section{Introduction}
Achieving a quantum advantage involves a delicate competition between entanglement and decoherence.
Recent experiments demonstrate the potential quantum advantage in random circuit sampling~\cite{Arute2019, Zhong2020, Morvan2024}, where noisy intermediate-scale quantum (NISQ) devices~\cite{Preskill2018, AbuGhanem2024b} can generate highly entangled quantum states before they are destroyed by noise.
Meanwhile, classical simulation using tensor network (TN) methods~\cite{Orus2014, Cirac2021} consumes significant computing resources~\cite{Pan2022A, Pan2022B}.
However, realizing the practical quantum advantage and solving valuable problems in physics or other fields beyond the capabilities of classical computers remains a much more challenging task~\cite{Daley2022}.
The level of difficulty varies in different contexts.
For instance, conventional quantum algorithms with theoretical promise, such as Shor's factoring or Grover's searching algorithms~\cite{Nielsen2009, AbuGhanem2025a}, require fault-tolerant quantum computers due to their reliance on non-local gates.
On the other hand, heuristic quantum-classical hybrid algorithms, such as the variational quantum eigensolver~\cite{Endo2021, Cerezo2021}, face scalability challenges due to barren plateaus, which lead to exponentially increasing measurement costs when training parameterized quantum circuits~\cite{Cerezo2024, Larocca2024}.

A recent experiment demonstrated quantum utility in simulating the time evolution of the Ising model using a discrete Trotter step of $\tau=\pi/2$, where the corresponding two-qubit gate can be directly compiled into a single CNOT gate along with other single-qubit gates~\cite{Kim2023}.
This work represents a significant advance toward the practical application of quantum computers in addressing real-world problems, moving beyond random circuit sampling that has limited practical relevance.
However, it does not preclude the possibility that their experimental results could be reproduced by advanced classical algorithms, as demonstrated by two subsequent numerical studies~\cite{Liao2023, Tindall2024}.
Meanwhile, accurately simulating time evolution, a key aspect in quantum many-body physics, typically involves much smaller Trotter steps.
These smaller steps result in deeper quantum circuits that are more vulnerable to decoherence~\cite{Guo2023}. 
This creates a trade-off between the compilation errors introduced by the discrete Trotter step and the circuit errors caused by noise.
From another perspective, the accumulation of quantum entanglement in a Trotter circuit with small $\tau$ progresses more slowly than in random quantum circuits, meaning that the entangling gates are not being fully utilized.

Fortunately, recent research~\cite{Luchnikov2021, McKeever2023, McKeever2024, Gibbs2024, Rogerson2024, Zemlevskiy2024} has shown that shallow quantum circuits can be used to approximate target quantum states or unitaries using various classical optimization techniques, especially combining with TN methods.
However, the optimized circuits in these studies still involve arbitrary two-qubit gates that are not directly implementable on quantum hardware.
In standard compilation strategies for superconducting quantum computers, a two-qubit gate is decomposed into three CNOT gates~\cite{Vidal2004}.
This increases both the classical optimization to train two-qubit gates and the quantum compilation costs, making this process more resource-intensive.
Moreover, few studies account for circuit noise, a significant constraint on the performance of state-of-the-art quantum devices.

In this paper, we propose a new approach to address these challenges by approximately compiling target quantum circuits while maximizing the overall fidelity, considering both compilation errors and noise effects.
Our key innovation is the direct optimization of a brick-wall circuit composed of trainable single-qubit gates and fixed two-qubit CNOT gates.
This eliminates the need for further compilation, allowing us to design a circuit that approximates the target quantum dynamics with high fidelity while maintaining shallow depth, making it suitable for accurate execution on a quantum computer.
Numerical simulations demonstrate that our method consistently outperforms the original circuit regarding overall fidelity in many cases, with reduced circuit noise compensating for approximation errors.
This is particularly evident in the time evolution of the critical Ising model and the quantum Fourier transformation (QFT) circuit, where the entanglement accumulates slowly at each step.
However, we observe little improvement for random circuits, which aligns with our intuitive expectations.
Experiments on IBM quantum computers demonstrate the effectiveness of our method in compiling quantum circuits, as exemplified by the time evolution of the critical Ising model with $N=12$ qubits, a time interval $\tau = 0.1$, and $200$ time steps compiled to an optimized circuit with depth $d_{\rm optim} = 48$.
Our approach is scalable to the number of qubits, paving the way for realizing a practical quantum advantage in near-term quantum devices before fault-tolerant quantum computing becomes feasible.

\section{Methods}
\begin{figure*}
    \centering
    \includegraphics[width=0.85\linewidth]{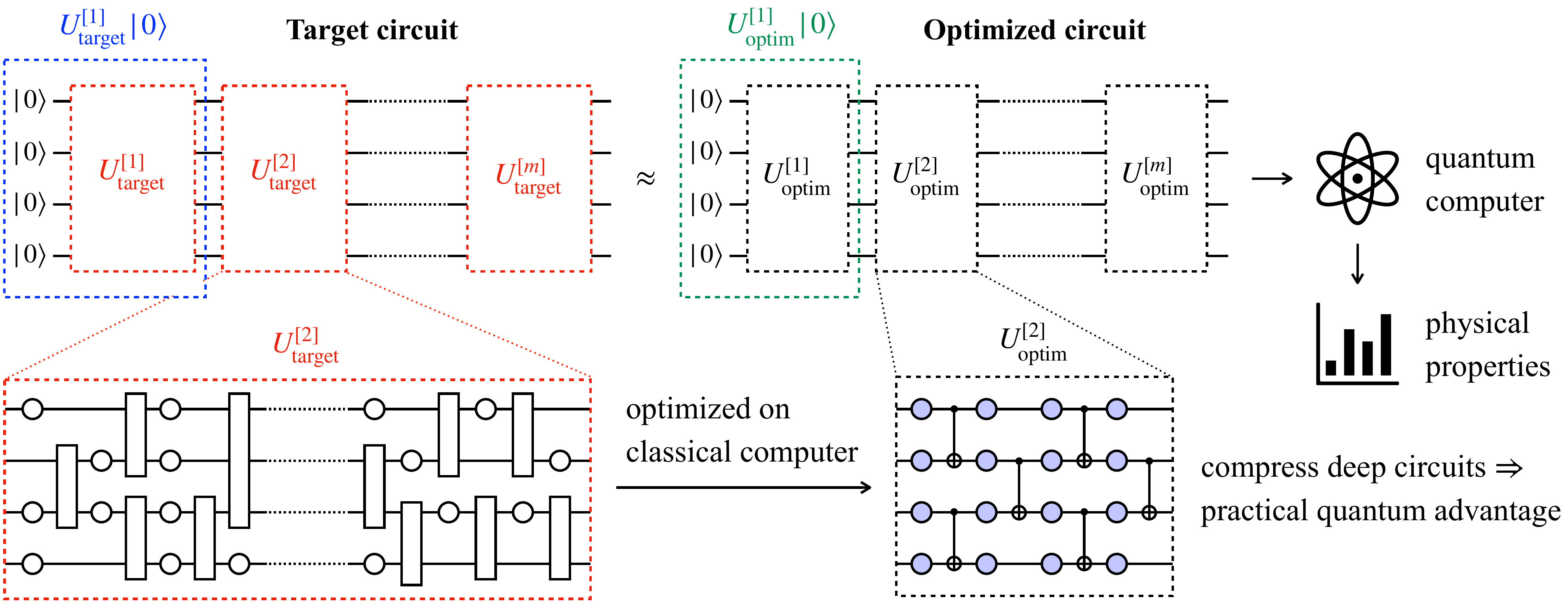}
    \caption{
    The overall framework of this study.
    A deep quantum circuit is divided into several parts, each with shallow depth enabling classical simulation.
    Each part is replaced by a brick-wall circuit composed of trainable single-qubit gates and fixed CNOT gates.
    The optimized circuit is subsequently executed on a quantum computer, where relevant observables are measured and collected.
    }\label{Fig: Framework}
\end{figure*}
\subsection{Approximate compilation of deep circuits}
Our method for compiling a target quantum circuit $U_{\rm target}$ involves the following key steps, as outlined in Fig.~\ref{Fig: Framework}.
\begin{enumerate}
    \item Break down the target circuit into manageable parts.
    \item Use a classical optimization algorithm to find an approximate compilation for each part through the following procedure
    \begin{itemize}
        \item Simulate the target circuit using a matrix product operator (MPO) representation.
        \item Initialize a brick-wall circuit structure, comprising trainable single-qubit gates and fixed CNOT two-qubit gates.
        \item Iteratively optimize single-qubit gates using the Riemannian optimization technique to maximize the fidelity between the optimized circuit and the target circuit on a classical computer.
    \end{itemize}
    \item Execute the optimized circuit on the quantum computer and measure relevant observables.
\end{enumerate}

Our compilation algorithm cannot be directly applied to a deep circuit that offers a potential quantum advantage, as it cannot be classically simulated and optimized.
Consequently, we assume the entire circuit consists of $m$ parts as
\begin{align}
    U_{\rm target} = U_{\rm target}^{[m]}\circ U_{\rm target}^{[m-1]}\circ \cdots \circ U_{\rm target}^{[1]},
\end{align}
where each part $U_{\rm target}^{[j]}$ has a fixed depth $d_{\rm target}^{[j]}$ enabling a classical representation as a MPO.

The second step is to optimize a reduced-depth quantum circuit for each part $U_{\rm target}^{[j]}$.
We assume that this optimized circuit has a brick-wall structure with trainable single-qubit gates and fixed CNOT two-qubit gates, as shown in Fig.~\ref{Fig: Framework}.
This structure can be directly executed on superconducting quantum computers without incurring additional costs.
Moreover, there is a minor adjustment for handling the first part $U_{\rm target}^{[1]}$, given that the input state $\ket{\psi_0}$ of the quantum circuit is known in advance.
Typically, this is set as a product state $\ket{0\cdots 0}$.
Otherwise, the preparing circuit for the initial state can be incorporated into the computational circuit.
For the problem we focus on, namely executing the compiled quantum circuits and obtaining measurement outcomes crucial for practical applications, the first part only needs to satisfy
\begin{align}
    U_{\rm optim}^{[1]}\ket{\psi_0} \approx U_{\rm target}^{[1]}\ket{\psi_0}.
\end{align}
For the remaining parts, we have
\begin{align}
    U_{\rm optim}^{[j]}\approx U_{\rm target}^{[j]},\quad j = 2, \cdots m
\end{align}

Specifically, each part of the target circuit is simulated by a matrix product state (MPS) or MPO for $U_{\rm target}^{[1]}\ket{\psi_0}$ or $U_{\rm target}^{[j]} (j = 2, \cdots, m)$, respectively, with the standard truncation method and a large enough bond dimension $D_{\rm target}$ to guarantee the accuracy (See Appendix).
This simulation is reasonable if we fix the maximum depth of each part as a constant $d_{\rm target}^{[j]}$ independent of the system size $N$.
The loss function is defined as follows
\begin{align}
    \Theta^{[1]} &= \left|U_{\rm optim}^{[1]}\ket{\psi_0} - U_{\rm target}^{[1]}\ket{\psi_0}\right|^2,\label{Equ: Loss1}\\
    \Theta^{[j]} &= \llvert U_{\rm optim}^{[j]} - U_{\rm target}^{[j]}\rrvert_F^2,\quad j = 2, \cdots m,\label{Equ: Loss2}
\end{align}
where $\llvert\cdot\rrvert_F$ denotes the Frobenius norm of an operator.
We optimize single-qubit gates using the Riemannian optimization technique to ensure the preservation of the unitary condition~\cite{Hauru2021, Luchnikov2021, Wiersema2023, Rogerson2024}, combined with state-of-the-art automatic differentiation (AD)~\cite{Rumelhart1986, Novikov2021} and the Adam optimizer~\cite{Kingma2017, Becigneul2019, Brantner2024}.

\subsection{Optimization of single-qubit gates}
Riemannian optimization is a powerful framework for addressing optimization problems subject to smooth manifold constraints~\cite{Udriste2013, Smith2014}.
By accounting for the intrinsic curvature of the underlying manifold, it enables the development of efficient, structure-preserving algorithms through the projection of gradients onto the tangent space of the manifold.
This approach has found successful applications across a wide range of disciplines, including machine learning, signal processing, computer vision, and quantum physics.
Its ability to naturally accommodate constraints such as orthogonality, low rank, and fixed norm makes it especially well-suited for modern large-scale and structured optimization problems.

In our case, we employ the Riemann optimization on the unitary manifold to determine the single-qubit gates that minimize the loss functions in Eqs.~\eqref{Equ: Loss1} and \eqref{Equ: Loss2}.
Specifically, in each iteration step, we calculate the gradient $\nabla_i$ for a gate (tensor) $U_i$ using AD, a computational technique that efficiently and accurately computes derivatives of functions by systematically applying the chain rule to elementary operations.
In particular, important matrix functions such as singular value decomposition (SVD) are compatible with AD for general complex matrices~\cite{Wan2019}.
The gradient is then projected onto the tangent space of the unitary manifold, 
\begin{align}
    \nabla_i^{\textrm{P}} = \frac{1}{2}\left(U_i^{\dagger}\nabla_i - \nabla_i^{\dagger}U_i\right).
\end{align}
The gate is then updated according to
\begin{align}
    U_i\Rightarrow U_ie^{-\eta \nabla_i^{\textrm{P}}},
\end{align}
where $\eta$ is the current learning rate automatically adjusted by the Adam optimizer, whose hyperparameters take the conventional values $\eta^0 = 1\times 10^{-3}$, $\beta_1 = 0.9$, $\beta_2 = 0.999$, and $\epsilon = 10^{-8}$.

\subsection{Metrics for evaluating the performance}
The fidelity between two normalized pure states is defined as
\begin{equation}
    f\left(\ket{\psi}, \ket{\phi}\right) = \left|\braket{\psi|\phi}\right|^2.
    \label{equ: fidelity}
\end{equation}
In our study, this definition is generalized for two unitaries (operators) as
\begin{equation}
    f\left(U_1, U_2\right) = \frac{1}{2^N}\left|\Tr{\left[U_1^{\dagger}U_2\right]}\right|,
\end{equation}
corresponding to the inner product in the operator space.

In the following sections, the compression rate is defined as
\begin{align}
    \gamma^{[j]} = \frac{\textrm{$n_{\rm CNOT}$ to compile $U_{\rm target}^{[j]}$}}{\textrm{$n_{\rm CNOT}$ in $U_{\rm optim}^{[j]}$}},
\end{align}
which evaluates the performance of our approximate compilation method compared to the standard one, focusing on the reduction of experimental cost and circuit noise.
For a target circuit that also has a brick-wall circuit but with general two-qubit gates (such as time evolution of nearest-neighbor quantum models or commonly studied random circuits), the above expression simplifies to
\begin{align}
    \gamma^{[j]} = \frac{3d_{\rm target}}{d_{\rm optim}},\label{Eq: compression}
\end{align}
where we utilize the fact that any two-qubit gate can be compiled by three CNOT gates.

\section{Results}
\subsection{Numerical results for critical Ising model}
\begin{figure*}
    \centering
    \includegraphics[width=0.67\linewidth]{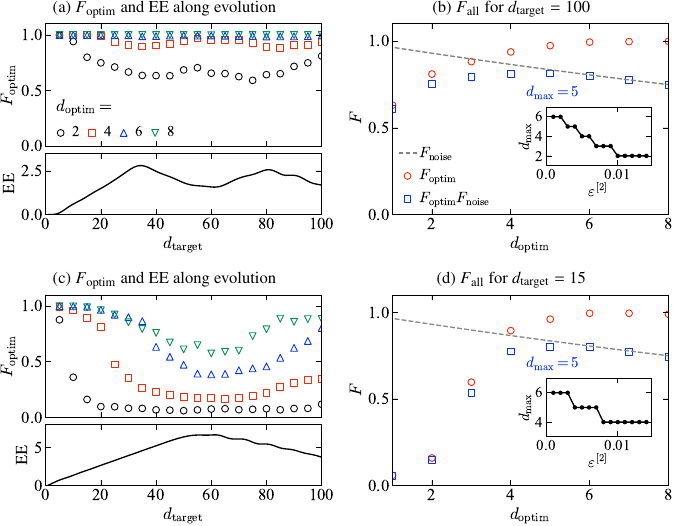}
    \caption{
    Performance of circuit compilation for the time evolution of critical Ising model with $N=10$ and $\tau=0.1$, simulated with $D_{\rm target}=128$.
    (a,b) Results for the output state from an initial product state. 
    (c,d) Results for the entire unitary.
    (a, c) Optimization fidelity $F_{\rm optim}$ for different $d_{\rm optim}$ and EE along the evolution.
    (b, d) Overall fidelity $F_{\rm all}$ for the state with $d_{\rm target}=100$ and unitary with $d_{\rm target}=15$ for a typical error rate $\varepsilon^{[2]}=4\times 10^{-3}$. 
    Insets show the optimal depth $d_{\rm max}$ for different error rates $\varepsilon^{[2]}$.
    }\label{Fig: Ising}
\end{figure*}

In this section, we present numerical results for the time evolution $e^{-i H t}$ of the critical Ising model with $N=10$, whose Hamiltonian is given by
\begin{align}
    H = -\sum_{i}\sigma_{i}^{z}\sigma_{i+1}^{z} - \sum \sigma_{i}^x.
\end{align}
We choose the first-order Trotter circuit with a time step $\tau = 0.1$ and total evolution time $t=10$ as the target circuit $U_{\rm target} = \left(e^{-iH\tau }\right)^{d_{\rm target}}$ with $d_{\rm target}=100$.
The target circuit with or without an initial product state is simulated using MPS or MPO with $D_{\rm target}=128$.
After that, an optimized circuit is used to approximate the target circuit, with optimization fidelity $F_{\rm optim}$ shown in Fig.~\ref{Fig: Ising}(a) and \ref{Fig: Ising}(d), respectively.
Furthermore, we display the entanglement entropy (EE) of the quantum state or operator during time evolution, with the logarithm taken to the base $2$.

Overall, the optimization fidelity is relatively higher when the initial state is fixed, which is as expected since the Hilbert space to be explored is much smaller in this case (about the square root of the entire unitary).
From the perspective of quantum entanglement, the accumulation of EE is faster in the full unitary than in the time-evolving state (comparing the lower panels of Fig.~\ref{Fig: Ising}(a) and \ref{Fig: Ising}(c)).
In both cases, the entanglement capacity of the CNOT gate is bounded by a constant value of $1$ for both state EE and operator EE.
Specifically, preparing a thermalized quantum state with $N$ qubits requires at least $N/2$ layers of CNOT gates, aligned with the observation in Fig.~\ref{Fig: Ising}(a) that $d_{\rm optim}=5$ is sufficient for any target circuit depth $d_{\rm target}$. 
However, to approximate a unitary for long-time evolution, at least twice the depth (i.e., $N$ layers of CNOT gates) is necessary.
Therefore, a higher compression rate can be achieved for the first part of the circuit than for the rest, i.e., $\gamma^{[1]}>\gamma^{[j]}$ for $j=2, \cdots, m$.
The above discussion implies a strong relation between $F_{\rm optim}$ and EE for a given $d_{\rm optim}$, which will be further discussed later.

Nevertheless, real quantum devices in the NISQ era are affected by noise, requiring a comprehensive consideration of various error sources.
To estimate the overall fidelity, we define $F_{\rm all} \equiv F_{\rm optim}F_{\rm noise}$, where $F_{\rm noise}$ accounts for the noise effects on the final performance.
We model this term as $F_{\rm noise} = \left(1-\varepsilon^{[2]}\right)^{n_{\rm CNOT}}$, where $\varepsilon^{[2]}$ represents the error rate for a two-qubit CNOT gate and $n_{\rm CNOT}=(N-1) d_{\rm optim}$ is the number of CNOT gates in the optimized brick-wall circuit.
We first consider a typical $\varepsilon^{[2]}=4\times 10^{-3}$ based on the latest advances in superconducting quantum computers~\cite{Acharya2024}.
The curves in Fig.~\ref{Fig: Ising}(b) and \ref{Fig: Ising}(d), corresponding to the state with $d_{\rm target}=100$ and unitary with $d_{\rm target}=20$, respectively, show a competition between the compilation error and noise effects.
Subsequently, we define an optimal depth $d_{\rm max}$ that maximizes the overall fidelity, and compare $d_{\rm max}$ across different values of $\varepsilon^{[2]}$ ranging from $1\times 10^{-3}$ to $1.5\times 10^{-2}$, covering the near-term performance of various platforms, as shown in the insets.
Combining these results with the inverse relation between $\gamma $ and $d_{\rm optim}$ in Eq.~\eqref{Eq: compression}, we conclude that stronger circuit noise enables a larger compression rate, reflecting a trade-off between the compilation error and noise effects.
Even for $\varepsilon^{[2]}=1\times 10^{-3}$, which is significantly lower than the noise levels of the most advanced quantum hardware, we still achieve compression of the circuit depth by nearly an order of magnitude, as demonstrated by the results in Fig.~\ref{Fig: Ising}(d) ($\gamma = 3\times 15/6 = 7.5$).

\subsection{Scalability to larger systems}
\begin{figure*}
    \centering
    \includegraphics[width=\linewidth]{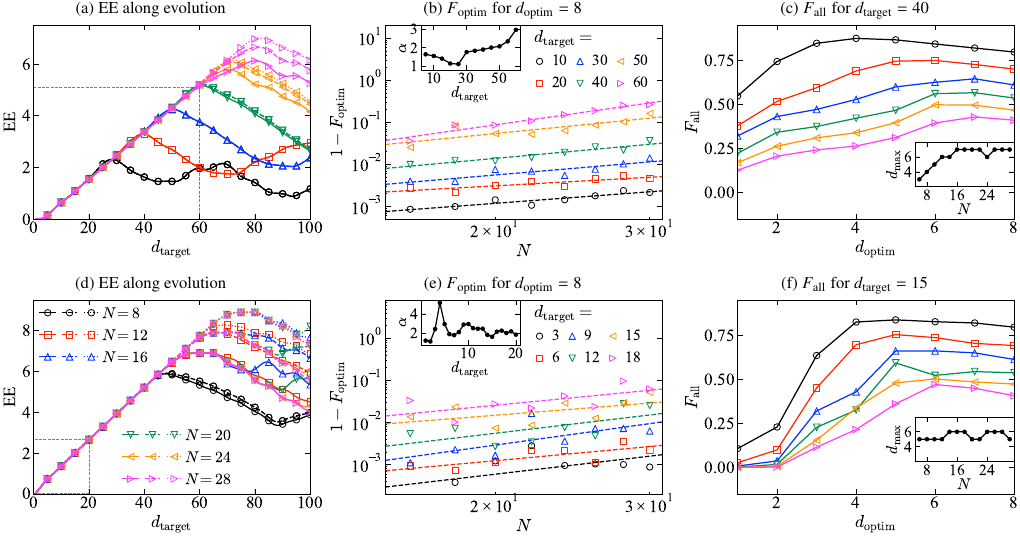}
    \caption{
    The scalability of circuit compilation for the time evolution of critical Ising model with $\tau=0.1$.
    (a-c) Results for the output state from an initial product state.
    (d-f) Results for the entire unitary.
    (a, d) EE dynamics for different $D_{\rm target}$ and $N$ (solid lines for $D_{\rm target}=128$, dashed lines for $D_{\rm target}=256$, and dotted lines for $D_{\rm target}=512$.
    Grey dashed frames mark the regimes suitable for efficient classical simulation and optimization.
    (b, e) Compilation error against increasing $N$ on a log-log scale for different $d_{\rm target}$ and $D_{\rm target}=128$.
    Insets show the fitted exponent $\alpha$ for $1-F_{\rm optim}\sim N^{\alpha}$.
    (c, f) Overall fidelity with varying $d_{\rm optim}$ for different $N$ and $D_{\rm target}=128$.
    Insets show the optimal depth $d_{\rm max}$.
    Note that (a), (c), (d), and (f) share a common legend plotted in (d).
    }\label{Fig: Scaling}
\end{figure*}

In this section, we explore the scalability of our scheme with respect to system size.
We begin by simulating the EE dynamics in Fig.~\ref{Fig: Scaling}(a) and \ref{Fig: Scaling}(d) for systems up to $N=30$ qubits, with varying bond dimensions (solid lines for $D_{\rm target}=128$, dashed lines for $D_{\rm target}=256$, and dotted lines for $D_{\rm target}=512$).
The convergence of EE as the bond dimension $D_{\rm target}$ increases confirms that the simulation of the target circuit is accurate, which is essential for the subsequent circuit compilation process.
Correspondingly, to ensure that each part of the circuit remains shallow and implementable by classical simulation and optimization, we bound the target circuit depths by constants $d_{\rm optim}^{[1]}=60$ and $d_{\rm optim}^{[j]}=40$ for $j=2, \cdots, m$, independent of system size.
Meanwhile, the results in Fig.~\ref{Fig: Ising}(d) suggest that compiling a unitary with $d_{\rm target}=20$ to an optimized circuit of $d_{\rm optim}=8$ yields satisfactory performance (with $F_{\rm optim}=0.963$).
Thus, we limit our analysis to shallow circuits with depths smaller than $d_{\rm optim}^{[j]}=20$ for $j=2, \cdots, m$ while still achieving a large compression rate of $\gamma=7.5$ that significantly reduces the noise effects typically associated with deeper circuits.

The final accuracy of our scheme is determined by the compilation fidelity of each part, especially the scaling law with a given constant depth but increasing $N$.
Specifically, we compare the trend of $1-F_{\rm optim}$ with increasing $N$ for a fixed optimized circuit depth $d_{\rm optim}=8$ and different target circuit depths up to $d_{\rm target}=60$ for the state in Fig.~\ref{Fig: Scaling}(b) and up to $d_{\rm target}=20$ for the unitary in Fig.~\ref{Fig: Scaling}(e), respectively, on a log-log scale.
The linearly fitted slopes in the insets reveal a power law of $1-F_{\rm optim}\sim N^{\alpha}$, with the exponent $\alpha$ fitted in the insets depending on the specific circuit depth $d_{\rm target}$.
Since our algorithm does not require scalability with increasing $d_{\rm target}$ (which would be impossible), a small exponent around $2$ for the target depth involved here is sufficient to demonstrate the efficiency and scalability of our algorithm for larger system sizes.

Next, we consider the scaling of quantum resources when accounting for circuit noise.
The overall fidelity $F_{\rm all}$ and the optimal depth $d_{\rm max}$ for the state at $d_{\rm target}=40$ and the unitary at $d_{\rm target}=15$ are depicted in Fig.~\ref{Fig: Scaling}(c) and \ref{Fig: Scaling}(f) for $\varepsilon^{[2]}=4\times 10^{-3}$, respectively.
The convergence of $d_{\rm max}$ with increasing system size $N$ demonstrates that only a constant depth (and thus a linear dependence $O(N)$ for the number of gates) is required to achieve the highest overall fidelity.
This behavior is related to previous numerical studies, which show that a fixed error rate will lead to decoherence of the entire system after a constant circuit depth, regardless of the system size~\cite{Noh2020, Guo2024}.

\subsection{Real experiments on IBM quantum devices}
\begin{figure}
    \centering
    \includegraphics[width=\linewidth]{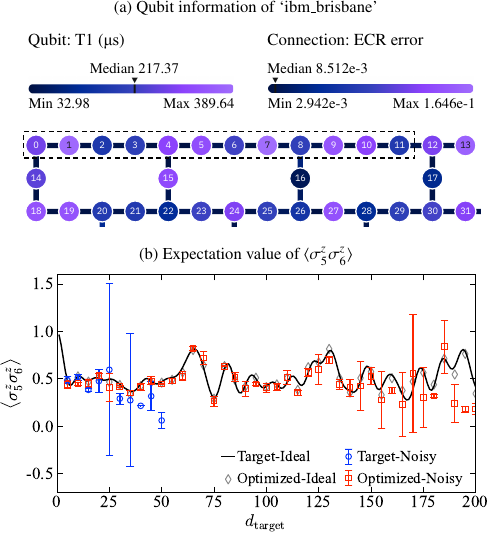}
    \caption{
    Real experiments on IBM quantum hardware.
    (a) Hardware information for `ibm\_brisbane'.
    Qubits $0\sim 11$ are used in our experiments.
    (b) Expectation values of $\sigma_5^z\sigma_6^z$ compared for ideal and noisy, target and optimized circuits.
    Errorbars indicate the standard deviation for experiment results.
    }\label{Fig: IBM}
\end{figure}

In this section, we verify our compilation scheme with real experiments.
Specifically, we present experimental results of executing optimized circuits on the IBM superconducting quantum computer `ibm\_brisbane' to simulate the time evolution of the critical Ising model with $N=12$, $\tau=0.1$, and $t=20$~\cite{AbuGhanem2025}.
The hardware details are depicted in Fig.~\ref{Fig: IBM}(a), where we use qubits $0\sim11$ for our experiments.
In this quantum processor, the native two-qubit entangling gate is the echoed cross-resonance (ECR) gate, which is just the entangling part of the CNOT gate~\cite{ECR}.
In other words, transpilation from our CNOT brick-wall circuits to an ECR-based one does not require additional quantum resources concerning the number of two-qubit gates.
The target circuit is divided into $m=6$ parts, with $d_{\rm target}^{[1]}=100$, $d_{\rm target}^{[j]}=20$ for $j=2, \cdots, 6$, and $d_{\rm optim}^{[j]}=8$ for $j=1, \cdots, 6$, corresponding to a total compression rate of $\gamma=3\times 200/48=12.5$.
The results comparing the ideal target circuit, the experimental target circuit, the ideal optimized circuit, and the experimental optimized circuit are shown in Fig.~\ref{Fig: IBM}(b), where the dynamics of a local observable $\sigma_5^z\sigma_6^z$ are used to evaluate the performance.
In the legend, ``Target'' and ``Optimized'' refer to the original time-evolution circuit and our optimized circuit, while ``Ideal'' and ``Noisy'' represent the results obtained from exact classical simulations and experiments on real quantum hardware (with standard deviation).
Error mitigation techniques, including dynamical decoupling (DD)~\cite{Viola1998}, Pauli twirling (also known as randomized compiling, RC)~\cite{Wallman2016}, twirled readout error extinction (TREX)~\cite{Berg2022}, and zero noise extrapolation (ZNE)~\cite{Temme2017}, are integrated into the experiments with the following specific settings and parameters.
DD is implemented by inserting pulse sequences $XX$ to suppress coherent errors on idle qubits.
RC mitigates coherent errors by inserting random Pauli gates, thereby converting them into Pauli stochastic errors.
We sample $64$ circuit instances from the ensemble of twirled circuits.
TREX corrects readout errors by learning the readout error transfer matrix by twirling the measurement over $32$ random circuits.
ZNE amplifies the noise strength by factors of ${1, 3, 5}$ and employs exponential extrapolation to estimate the zero-noise limit.

Direct execution of the target circuit (blue circles) is hindered by decoherence, limiting it to depths of $d_{\rm target}=50$ and large uncertainties nearby.
Moreover, circuits exceeding depths $d_{\rm target}=20$ cannot provide reliable results, highlighting the current limitations of quantum hardware in executing Trotter circuits for Hamiltonian evolution.
In contrast, the implementation of our optimized circuit (red squares) enables effective simulation of the target dynamics with a small error bar up to $d_{\rm target}=160$ (with $d_{\rm optim}=32$), significantly extending the range of quantum simulation possible with NISQ quantum hardware by one order of magnitude.
This result demonstrates the potential of approximate circuit compilation in overcoming the limitations of quantum hardware, especially in simulating more complex systems.
We also extend our investigations to other observables, yielding similar results (not shown here).
Furthermore, the comparison between ideal and experimental data for the optimized circuit in Fig.~\ref{Fig: IBM}(b) reveals an important observation. 
The distance between gray diamonds and the black curve is much smaller than the distance between gray diamonds and red squares, especially for large $d_{\rm target}$.
This suggests that noise effects in the optimized circuit still dominate over compilation errors, implying that further improvement in compression rate could be achieved.
However, the trade-off between noise and compilation error is hardware-dependent, determined by the specific error rates of the quantum device as demonstrated in the previous section.
Therefore, an optimal depth should be carefully chosen for each specific task to maximize fidelity and fully leverage the potential of quantum computers.

\subsection{Quantum Fourier transformation}
\begin{figure*}
    \includegraphics[width=\linewidth]{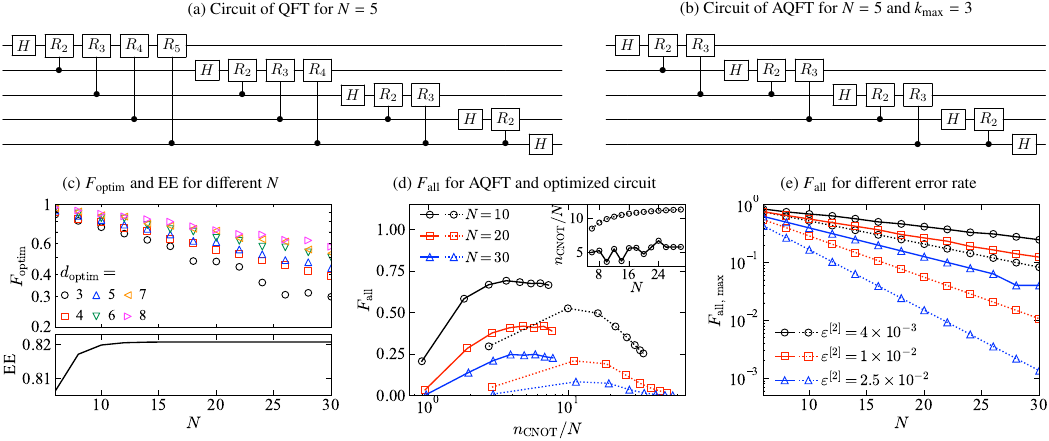}
    \caption{
    Results for QFT without SWAP gates at the end.
    (a) Quantum circuit of QFT without the final SWAP gates.
    (b) Quantum circuit of AQFT for $k_{\rm max}=3$ without the final SWAP gates.
    (c) Optimization fidelity $F_{\rm optim}$ and EE for different $N$ and $d_{\rm optim}$.
    (d) Overall fidelity $F_{\rm all}$ of the optimized circuit (solid lines) and AQFT circuit (dotted lines) against the number of CNOT gates used.
    Insets compare the optimal $n_{\rm CNOT}/N$ for both circuits across different $N$.
    (e) Maximum of overall fidelity $F_{\rm all,\ max}$ of the optimized circuit (solid lines) and AQFT circuit (dotted lines) at different error rates.
    }\label{Fig: QFT}
\end{figure*}

In the following, we explore the performance of our compilation method in other tasks beyond simulating time evolution.
Here we consider the QFT circuit, which plays a central role in many important and promising quantum algorithms~\cite{Nielsen2009}.
The QFT circuit $F_N$ can be decomposed as $F_N = S_NQ_N$, where $Q_N$ is the core part that performs the Fourier transformation of the information encoded in the input state, as shown in Fig.~\ref{Fig: QFT}(a).
$S_N$ is a sequence of SWAP gates that reverse the order of all qubits at the end of the circuit (not shown here).
The circuit $Q_N$ contains non-local control-rotation gates $CR_k$, where $k$ is the interaction length and $R_k$ represents the following rotation
\begin{align}
    R_k = \left[
    \begin{array}{cc}
    1 & 0\\
    0 & e^{i \frac{2\pi}{k}}
    \end{array}
    \right].
\end{align}
It has been analytically proved that $Q_N$ features a set of exponentially decaying Schmidt coefficients independent of the system size $N$~\cite{Chen2023}.
This insight motivates us to investigate a more compact version of this circuit, as the original configuration is quite sparse.
Fig.~\ref{Fig: QFT}(c) illustrates the optimization fidelity for the entire $Q_N$ as $N$ increases on a logarithmic scale, accompanied by the operator EE that clearly converges to a finite value as $N$ grows.
The exponential decay of $F_{\rm optim}$ with $N$ is reasonable, since the circuit depth of $Q_N$ scales quadratically with $N$, rather than being fixed.

Similar to the previous analysis, we also consider the impact of noise effects by assuming a fixed error rate $\varepsilon^{[2]}=4\times 10^{-3}$ for the CNOT gates.
The overall fidelity $F_{\rm all}$ and the corresponding number of CNOT gates $n_{\rm CNOT} = (N-1)d_{\rm optim}$ (which quantifies the quantum resources required to execute the circuit) are depicted in Fig.~\ref{Fig: QFT}(d) by solid lines.
For comparison, we evaluate the performance of approximate QFT (AQFT)~\cite{Barenco1996} shown in Fig.~\ref{Fig: QFT}(b), where qubit rotations below a certain angle $2\pi/k_{\rm max}$ are neglected, as illustrated in Fig.~\ref{Fig: QFT}(d) by dotted lines.
The number of CNOT gates used to compile the AQFT circuit is derived in Methods.
The results indicate that our optimized circuit achieves higher accuracy than the AQFT circuit with the same amount of quantum resources.
Meanwhile, the optimal $n_{\rm CNOT}$ to achieve maximal overall fidelity $F_{\rm all}$ compared in the inset of Fig.~\ref{Fig: QFT}(d) demonstrate the reduction of quantum resources using our method.
In addition, we compare the maximal overall fidelities $F_{\rm all, max}$ of both circuits for three typical values of error rates ($\varepsilon^{[2]} = 4\times 10^{-3}, 1\times 10^{-2}, \textrm{and }2.5\times 10^{-2}$) in Fig.~\ref{Fig: QFT}(e), representing different stages of NISQ quantum hardware before the realization of fault-tolerant systems~\cite{AbuGhanem2024}.
The results for the optimized circuit consistently show an advantage over the AQFT circuit across various noise levels.
This advantage diminishes as the error rate decreases, suggesting reduced compressibility for quantum devices with better performance.

\subsection{Haar Random quantum circuits}
\begin{figure*}
    \centering
    \includegraphics[width=0.66\linewidth]{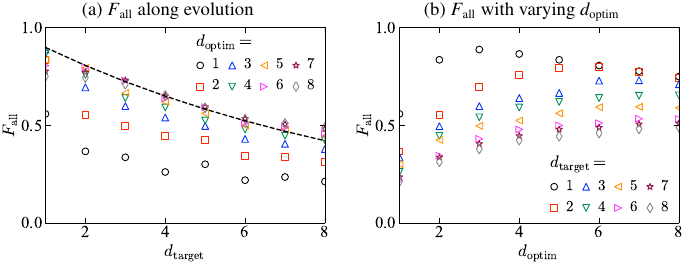}
    \caption{
    Performance of circuit compilation for Haar random quantum circuits with $N=10$.
    (a) Overall fidelity $F_{\rm all}$ with increasing target circuit depth $d_{\rm target}$ for different $d_{\rm optim}$.
    The black dashed line represents the direct compilation of the target circuit.
    (b) Overall fidelity $F_{\rm all}$ with varying $d_{\rm optim}$ for different $d_{\rm target}$.
    }\label{Fig: Random}
\end{figure*}

We evaluate the performance of our method by approximately compiling a Haar random quantum circuit and analyzing its characteristics.
A Haar random circuit exhibits a brick-wall layout, where each two-qubit gate is drawn randomly and independently from the uniform distribution in the unitary group $U(4\times4)$~\cite{Fisher2023}.
Such circuits are characterized by ballistic growth of entanglement, which complicates their classical simulation using the TN methods and is often associated with quantum advantage~\cite{Arute2019}.
The comparison in Fig.~\ref{Fig: Random}(a) for $N=10$ between direct compilation (black dashed line) and our approximate compilation method shows that little improvement can be achieved with our approach.
Meanwhile, the optimal depth $d_{\rm max}$ identified in Fig.~\ref{Fig: Random}(b) suggests that the random circuit can be hardly compressed (with the compression rate $\gamma=1$ for $d_{\rm target}=1,\ 2$).
This limitation can be attributed to the fact that the target circuit quickly accumulates entanglement, which maximizes the entangling capacity of the two-qubit gates, leaving little room for further performance improvement.

\section{Discussion}
In this study, we propose an alternative procedure for executing quantum circuits by approximately compiling them into a brick-wall layout, where CNOT gates are the only two-qubit gates involved.
In this section, we further discuss the advantage of our scheme over other compilation methods and its broad implications in various tasks and platforms.

\subsection{Relation with other circuit compilation methods}
It is worth emphasizing that our approximate compilation strategy is conceptually distinct from existing methods such as variational compiling~\cite{Sharma2020, Xu2021} and hardware-aware transpilation~\cite{Niu2020, Du2024}.
Variational compiling typically aims to reproduce a target unitary by variationally optimizing parameters within a hybrid quantum-classical framework similar to VQE, where parameters are updated based on measurement outcomes.
This process often necessitates deep circuits when the target operation possesses nontrivial entanglement structures, leading to significant noise accumulation in near-term quantum devices and exacerbating the barren plateau problem~\cite{Larocca2024}.
Hardware-aware transpilers, on the other hand, optimize the circuit mapping according to hardware connectivity and native gate sets, but still attempt to exactly implement the original quantum operation without introducing an approximation.
In contrast, our method allows for a controlled approximation of the target circuit, enabling substantial depth compression by optimizing a fixed brick-wall layout composed of CNOT and trainable single-qubit gates.
By sacrificing exact reproduction in favor of noise resilience, our approach improves the overall fidelity of practical computations on NISQ devices and advances the realization of practical quantum advantage.

\subsection{Transpilation to specific quantum hardware}
\begin{figure*}
    \includegraphics[width=\linewidth]{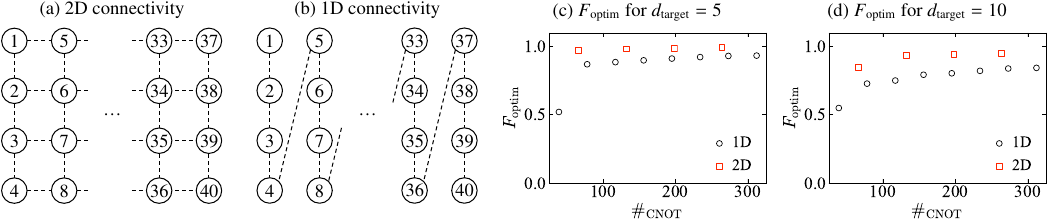}
    \caption{Performance of circuit compilation for 2D critical Ising time evolution on a square lattice with $N=4\times 10$ and $\tau=0.1$.
    (a) The qubit topology for 2D circuits.
    (b) The qubit topology for 1D circuits.
    The dashed line presents the connectivity structure.
    (c) Optimization fidelity for $d_{\rm target}=5$.
    (d) Optimization fidelity for $d_{\rm target}=10$.}
    \label{Fig: 2D}
\end{figure*}

In our method, we assume a brick-wall circuit structure defined on a one-dimensional (1D) spin array, with CNOT gates serving as native two-qubit gates.
We choose the CNOT gate as the basic element of the compiled circuits for two primary reasons.
Firstly, the CNOT gate is a maximally entangling gate that, when combined with other single-qubit gates, enables universal quantum computation.
Second, the CNOT gate is equivalent (up to single-qubit rotations) to the native entangling gates employed in a variety of quantum processors, including the CZ gate used in IBM's Heron and Google's Willow processors~\cite{Acharya2024}, as well as in neutral-atom quantum computers~\cite{Evered2023}; the ECR gate utilized in IBM's Eagle processor~\cite{ECR}; and the Mølmer–Sørensen (MS) gate in ion-trap systems~\cite{Soerensen1999}.
Consequently, our optimized circuits can be directly transpiled onto these platforms by accounting for specific hardware details, without incurring additional classical or quantum overhead.

On the other hand, recent progress in quantum hardware has demonstrated two-dimensional (2D) arrays on various platforms~\cite{Bluvstein2024, Guo2024A}.
Here, we present preliminary results for generalizing our compilation method to 2D systems.
Specifically, we consider the time evolution circuit of the critical Ising model with $\tau=0.1$ and given initial product state on a 2D square lattice of size $N_1\times N_2 = 4\times 10$, as shown in Fig.~\ref{Fig: 2D}(a).
To implement our scheme on this target circuit, we compare two types of qubit topology as accessible quantum hardware for the optimized circuit, including a 1D spin chain to simulate the 2D dynamics via a snape-shape coding pattern shown in Fig.~\ref{Fig: 2D}(b), as well as a 2D quantum system with the same structure as the target circuit shown in Fig.~\ref{Fig: 2D}(a).
Specifically, for a 2D optimized circuit, the number of CNOT gates is $\#_{\rm CNOT}=(2N_1N_2-N_1-N_2)d_{\rm optim}$.
The optimization fidelity of different circuit topologies (1D or 2D) is compared in Fig.~\ref{Fig: 2D}(c, d) for $d_{\rm target}=5$ and $10$, respectively.
In general, a high fidelity and compression rate (e.g., $F_{\rm optim}=0.944$ for $d_{\rm target}=10$ and $d_{\rm optim}=3$ ($\#_{\rm CNOT}=198$), leading to $\gamma=3\times 10/3=10$) can be achieved when compiling the target quantum circuit using a 2D quantum hardware with the same qubit topology.
Meanwhile, the results also demonstrate a consistent advantage of 2D quantum hardware over 1D in approximating the target 2D quantum circuit given the same number of CNOT gates.
Therefore, our method can be generalized to 2D quantum hardware on a square lattice to effectively compress the circuit depth and enhance the overall fidelity.
Further optimization for a specific qubit topology, such as a triangle or honeycomb lattice, is left for future study.

\subsection{Implications in different tasks}
\begin{figure*}
    \includegraphics[width=0.66\linewidth]{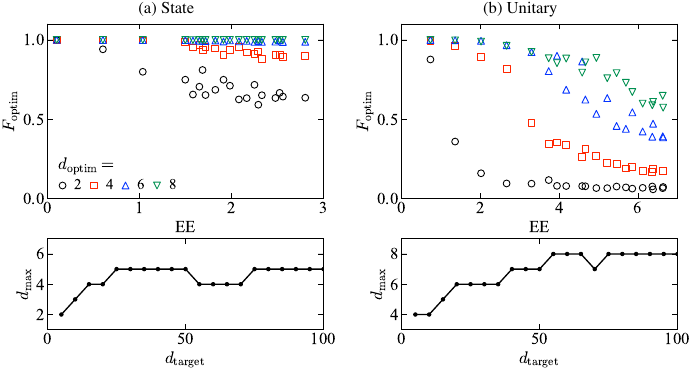}
    \caption{Upper panels: Relation between $F_{\rm optim}$ and EE for (a) state and (b) unitary in the time evolution of critical Ising model with $N=10$ and $\tau=0.1$.
    Lower panels: optimal depth $d_{\rm max}$ for the target circuits up to different depths $d_{\rm target}$ along the evolution.}
    \label{Fig: EE}
\end{figure*}

In general, our scheme can be integrated into various quantum computational tasks, enhancing overall accuracy and reducing experimental costs.
However, as previously discussed, the effectiveness of our approach depends on how quickly entanglement accumulates in the target circuit, limiting its applicability to certain types of random circuits that exhibit ballistic entanglement growth.
Table~\ref{Tab: Summary} summarizes the performance of our approximate circuit compilation method for different types of circuits considered in this study, with key indicators including the maximal overall fidelity $F_{\rm max}$, optimal circuit depth $d_{\rm max}$, and the compression rate with respect to the original circuit $\gamma$.
\begin{table}[h]
    \centering
    \begin{tabular}{c|c|c|c}\hline
        Circuit type & $F_{\rm max}$ & $d_{\rm max}$ & $\gamma$\\\hline
        Critical Ising time evolution at $d_{\rm target}=15$ & $0.802$ & $5$ & $9$ \\
        Quantum Fourier transformation & $0.692$ & $4$ & $9.75$ \\
        Haar random circuits at $d_{\rm target}=3$ & $0.730$ & $7$ & $1.29$\\\hline
    \end{tabular}
    \caption{Performance of circuit compilation for different circuits with $N=10$ and $\varepsilon^{[2]}=4\times 10^{-3}$.}
    \label{Tab: Summary}
\end{table}

In the following, we discuss the implications of our results in different contexts.
We first consider the relation between optimization fidelity and EE.
For the example of time evolution of the critical Ising model, this relation is explicitly shown in Fig.~\ref{Fig: EE}(a) and (b) for the approximation of state and unitary, respectively.
Moreover, the lower panels present the optimal depths $d_{\rm max}$ that for different stages throughout the evolution, following a trend similar to the corresponding EE dynamics of the target circuit shown in the insets of Fig.~\ref{Fig: Ising}(a, c).
In other words, whether our approximate compilation method can effectively realize depth compression is determined by the rate of entanglement accumulation.
In contrast, for random quantum circuits exhibiting a ballistic entanglement mambrane~\cite{Nahum2017}, any approximation in the compilation introduces a non-negligible error as demonstrated before, hindering the application of our scheme.

We note that while error mitigation techniques~\cite{Endo2021, Cai2023} are widely applied to improve the measurements of expectation values~\cite{Kim2023}, they are insufficient to overcome this limitation.
Random circuit sampling tasks focus on reproducing the full output distribution rather than estimating specific observables, making them particularly sensitive to fidelity loss induced by depth compression.
Moreover, the incompressibility of Haar-random circuits arises from their intrinsic requirement for deep, highly entangling dynamics, which cannot be fully captured by shallow approximations.
Although error mitigation could, in principle, be combined with both the original and compressed circuits, it cannot restore the lost entanglement or correct the fundamental approximation errors.
Therefore, the challenge presented by Haar-random circuits is intrinsic and cannot be readily addressed by hybrid strategies based on error mitigation.

The above discussion leads to the intriguing conclusion that only a constant depth is required to simulate time evolution with varying Trotter steps.
This comes from the fact that EE converges to a constant value when $t$ is fixed and $\tau$ decreases.
Consequently, we expect the compression rate to strongly depend on the Trotter step with $\gamma\sim 1/\tau$.
In this sense, simulating the target circuit using MPO simulation introduces only a classical overhead of $t/\tau$, while the quantum resource cost remains constant.
In particular, taking the Trotter step $\tau$ to an infinitesimally small value ($\tau \rightarrow 0$) would theoretically lead to an infinite compression rate ($\gamma \rightarrow \infty$), which may seem ideal. 
However, it does not imply that arbitrary precision can be achieved (which is also practically infeasible), as compilation errors and noise effects impose fundamental limits on attainable accuracy.
In this case, for sufficiently small values of $\tau$, the approximate compilation error dominates and the Trotter error becomes negligible.
Therefore, pursuing an infinitely small $\tau$ is unnecessary and may not yield substantial improvements.
For quantum computers in the NISQ era, the key challenge is balancing the trade-off between compilation error and the noise effects inherent in current quantum hardware.
Our scheme provides a pathway to help state-of-the-art NISQ devices transition from demonstrating quantum utility~\cite{Kim2023} to achieving a practical quantum advantage.
In the long term, reducing compilation error with increased quantum resources will not be a major focus until fault-tolerant quantum computers become available.

Another promising direction is to apply our scheme to promote the implementation of important quantum algorithms such as Shor's factoring and Grover's searching algorithms.
Our results for QFT indicate that a numerically optimized circuit can outperform an analytically constructed one in terms of approximately executing a quantum algorithm.
Recent studies have explored the classical simulation of these algorithms using TN methods~\cite{Chen2023, Niedermeier2024, Stoudenmire2024}, demonstrating that at least certain parts of these algorithms exhibit low entanglement or slow entanglement accumulation.
This implies that our scheme could be particularly effective in facilitating their fast and accurate execution.
In summary, our approach has the potential to foster the realization of powerful quantum algorithms on currently available quantum devices before fully fault-tolerant, enabling solutions to practical problems in various fields.

\section{Declarations}
\subsection{Availability of data and materials}
The datasets generated and analyzed during the current study are available from the corresponding author upon reasonable request.
The code for this study is available from the corresponding author upon reasonable request.
The IBM Quantum device `ibm\_brisbane’ is accessible at https://quantum-computing.ibm.com/.

\subsection{Competing interests}
The authors declare no competing interests.

\subsection{Funding}
This work is supported by the National Natural Science Foundation of China (NSFC) (Grant No. 12475022 and No. 12174214) and the Innovation Program for Quantum Science and Technology (Grant No. 2021ZD0302100).

\subsection{Authors' contributions}
Y. Guo conceived, designed, and performed the experiments.
Y. Guo and S. Yang analyzed the data and wrote the paper.
S. Yang contributed analysis tools and supervised the project.

\newpage

\renewcommand{\thesection}{S-\arabic{section}} \renewcommand{\theequation}{S%
\arabic{equation}} \setcounter{equation}{0} \renewcommand{\thetable}{S%
\arabic{table}} \setcounter{table}{0} \renewcommand{\thefigure}{S%
\arabic{figure}} \setcounter{figure}{0}
\section{Appendix}
\subsection{MPS and MPO for 1D systems}
\begin{figure*}
    \includegraphics[width=0.9\linewidth]{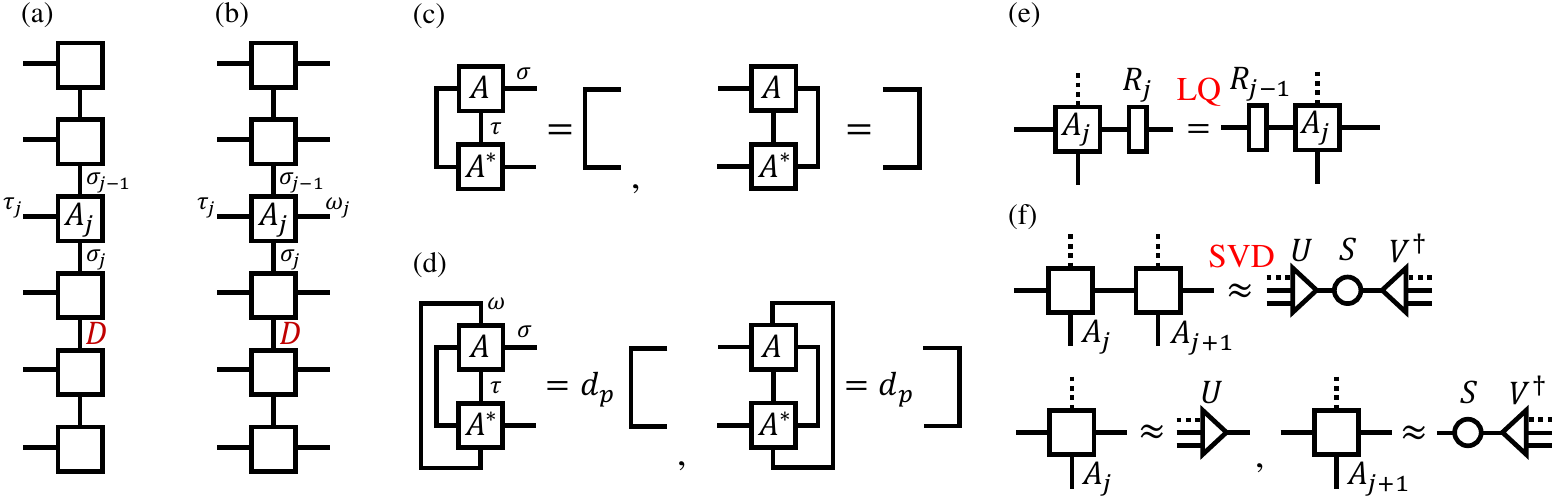}
    \caption{Truncation method for MPS and MPO.
    (a) MPS to present a quantum state.
    (b) MPO to present a quantum operator.
    (c) Cananical form for MPS.
    (d) Canonical form for MPO.
    (e) Right-to-left LQ decomposition to obtain the right-canonical form.
    (f) Left-to-right SVD to truncate each virtual index.
    }
    \label{Fig: TN}
\end{figure*}

Tensor network methods are powerful techniques used to efficiently represent and compute properties of complex quantum many-body systems, where direct methods would be computationally prohibitive due to the exponential growth of the Hilbert space~\cite{Verstraete2008, Orus2014, Cirac2021}.
The core idea behind these methods is to express a many-body quantum state as a network of smaller tensors connected by virtual indices, where each connection represents a summation over shared degrees of freedom.
This structure encodes entanglement and correlations in a compressed and systematic manner.
Tensor networks have found successful applications in various fields, including condensed matter physics, quantum information, and machine learning.
Their primary advantage lies in exploiting the entanglement structure of physical states, enabling otherwise intractable computations to become feasible.

Specifically, a 1D quantum state $\ket{\psi}$ with $N$ qubits satisfying the entanglement area law can be represented by an MPS,
\begin{align}
    \ket{\psi} = \sum_{\{\bm{\tau}\}}\sum_{\{\bm{\mu}\}}\prod_{j=1}^N[A_j]^{\tau_j}_{\mu_{j-1}, \mu_{j}}
    \ket{\tau_1,\cdots,\tau_N},
\end{align}
where $\bm{\tau}$ represent the physical degree of freedom with dimension $d_p$ ($d_p=2$ for spin systems), while $\bm{\mu}$ denote virtual indices with dimension $D$, as shown in Fig.~\ref{Fig: TN}(a).
Similarly, a 1D operator, e.g., a density matrix or a shallow quantum circuit $U$, can be represented by an MPO shown in Fig.~\ref{Fig: TN}(b)
\begin{align}
    U = \sum_{\{\bm{\tau}, \bm{\omega}\}}\sum_{\{\bm{\mu}\}}\prod_{j=1}^N[A_j]^{\tau_j, \omega_j}_{\mu_{j-1}, \mu_{j}}
    \ket{\tau_1,\cdots,\tau_N}\hspace{-0.5mm}\bra{\omega_1,\cdots,\omega_N},
\end{align}
where $\bm{\tau}$ and $\bm{\omega}$ correspond to the output and input physical indices, respectively.

In the simulation of quantum circuits, the bond dimension $D$ of the MPS or MPO must be truncated at each time step.
We utilize the canonical form of the MPS, as depicted in Fig. \ref{Fig: TN}(c), to eliminate the influence of the tensor environment during the compression of a virtual index. This naturally aligns with the normalization condition for the quantum state $\braket{\psi|\psi}=1$.
The canonical form of the MPO can be defined similarly, as shown in Fig. \ref{Fig: TN}(d), with an additional factor of $d_p$ to ensure the condition $\Tr{[U^{\dagger}U]} = d_p^N$.
After the implementation of each two-qubit gate layer, we employ a two-step process to obtain a compressed form for the current MPS or MPO~\cite{Orus2014}.
First, we implement a right-to-left LQ decomposition in Fig.~\ref{Fig: TN}(e) to obtain the right-canonical form, where $R_{N} = [1]$ is a $1\times 1$ identity matrix.
Second, we perform a left-to-right SVD to truncate the virtual bond to a constant value $D$, as illustrated in Fig.~\ref{Fig: TN}(f).

\subsection{Circuit of QFT}
QFT is the key ingredient of phase estimation, which constitutes many interesting quantum algorithms including order-finding and Shor factoring algorithms.
For a system with $N$ qubits, the QFT $F_N$ on an orthonormal basis $\ket{0}, \cdots, \ket{2^N-1}$ is defined as~\cite{Nielsen2009}
\begin{align}
    F_N\ket{j} = \frac{1}{2^{N/2}}\sum_{k=0}^{2^N-1}{e^{2\pi i jk/2^N}}\ket{k}.
\end{align}
Alternatively, QFT can be expressed in the binary representation,
\begin{align}
\begin{aligned}
    &F_N\ket{j_1, \cdots, j_N} = \frac{1}{2^{N/2}}\left(\ket{0}+e^{2\pi i 0.j_N}\ket{1}\right)\otimes\\
    &\left(\ket{0}+e^{2\pi i 0.j_{N-1}j_N}\ket{1}\right)\otimes\cdots\otimes\left(\ket{0}+e^{2\pi i 0.j_1\cdots j_N}\ket{1}\right),
\end{aligned}
\end{align}
where all decimals $0.j_l\cdots j_m$ are in binary representation.
The circuit shown in Fig.~\ref{Fig: QFT}(a) implements the QFT, but with a reversing order of qubits, i.e.,
\begin{align}
\begin{aligned}
    &Q_N\ket{j_1, \cdots, j_N} = \frac{1}{2^{N/2}}\left(\ket{0}+e^{2\pi i 0.j_1\cdots j_N}\ket{1}\right)\otimes\\
    &\left(\ket{0}+e^{2\pi i 0.j_2\cdots j_N}\ket{1}\right)\otimes\cdots\otimes\left(\ket{0}+e^{2\pi i 0.j_N}\ket{1}\right).
\end{aligned}
\end{align}
In the end, a set of SWAP gates $S_N$ is required to complete the QFT algorithm.

\subsection{Compilation of AQFT circuit}
When compiling the AQFT circuit on a real quantum device, the nonlocal gates $CR_k$ will introduce additional costs determined by the involvement of SWAP gates.
Fortunately, the rotation angle $2\pi/k$ in $R_k$ is inversely proportional to the interaction length $k$, suggesting a truncation of $k_{\rm max}$ to control the quantum resources.
This leads to the AQFT circuits shown in Fig.~\ref{Fig: QFT}(b).
Now we focus on the number of CNOT gates required to compile an AQFT circuit of $N$ qubits with a fixed $k_{\rm max}\leq N$.
The AQFT circuit naturally allows for a division into $N$ rounds, where the first $N-k_{\rm max} + 1$ rounds have the same structure, differing only by the translation of qubits.
The remaining $k_{\rm max}-1$ rounds exhibit decreasing interaction lengths.
Consider a round with maximum interaction length $k$ and starting qubit $j$.
To realize all nonlocal gates, we need to exchange the starting qubit with adjacent qubits to move it to site $k+j-2$, followed by a backmovement.
This results in a total number of $k-1 \text{ (gates) } + 2(k-2) \text{ (exchange) } = 3k-5$ two-qubit gates.
The interaction length and the corresponding number of two-qubit gates for each round are summarized in the following table.
\begin{table}[H]
    \centering
    \begin{tabular}{ccc}
    \hline
     Round No. & $k$ & Two-qubit gate number\\\hline
     $1$ & $k_{\rm max}$ & $3k_{\rm max}-5$\\
     &$\cdots$&\\
     $N-k_{\rm max}+1$ & $k_{\rm max}$ & $3k_{\rm max}-5$\\
     $N-k_{\rm max}+2$ & $k_{\rm max}-1$ & $3k_{\rm max}-8$\\
     &$\cdots$ &\\
     $N-1$ & $2$ & $1$\\
     $N$ & $1$ & $0$\\\hline
    \end{tabular}
\end{table}
Summing over all rounds and considering that each two-qubit gate is compiled into three CNOT gates, we obtain the total number of CNOT gates as
\begin{align}
\begin{aligned}
    n_{\rm CNOT} &= 3\left[(3k_{\rm max}-5)(N-k_{\rm max}+1) + \sum_{k=2}^{k_{\rm max}-1}(3k-5) \right] \\
    &= 3\left[(3k_{\rm max}-5)(N-\frac{k_{\rm max}}{2})-(k_{\rm max}-2)\right].
    \end{aligned}
\end{align}

\bibliography{ref}

\end{document}